\documentclass[rnote]{aa}

\usepackage{graphicx}
\usepackage{txfonts}
\usepackage{textcomp}
\usepackage{epstopdf}
\usepackage{subfigure}
\usepackage{natbib}
\usepackage[utf8]{inputenc}
\begin{document}

   \title{Why are Jupiter-family comets active and asteroids in cometary-like orbits inactive?}

   \subtitle{How hydrostatic compression leads to inactivity}

   \author{B. Gundlach\inst{1}, J. Blum\inst{1}}

   \institute{Technische Universit{\"a}t Braunschweig, Institut f{\"u}r Geophysik und extraterrestrische Physik, Mendelssohnstra{\ss}e 3, D-38106 Braunschweig, Germany\\
              \email{b.gundlach@tu-bs.de}
             }

   \date{Received , Accepted }


\abstract{Surveys in the visible and near-infrared spectral range have revealed the presence of low-albedo asteroids in cometary like orbits (ACOs). In contrast to Jupiter family comets (JFCs), ACOs are inactive, but possess similar orbital parameters.}
{In this work, we discuss why ACOs are inactive, whereas JFCs show gas-driven dust activity, although both belong to the same class of primitive solar system bodies.}
{We hypothesize that ACOs and JFCs have formed under the same physical conditions, namely by the gravitational collapse of ensembles of ice and dust aggregates. We use the memory effect of dust-aggregate layers under gravitational compression to discuss under which conditions the gas-driven dust activity of these bodies is possible.}
{Owing to their smaller sizes, JFCs can sustain gas-driven dust activity much longer than the bigger ACOs, whose sub-surface regions possess an increased tensile strength, due to gravitational compression of the material. The increased tensile strength leads to the passivation against dust activity after a relatively short time of activity.}
{The gravitational-collapse model of the formation of planetesimals, together with the gravitational compression of the sub-surface material simultaneously, explains the inactivity of ACOs and the gas-driven dust activity of JFCs. Their initially larger sizes means that ACOs possess a higher tensile strength of their sub-surface material, which leads to a faster termination of gas-driven dust activity. Most objects with radii larger than $2 \, \mathrm{km}$ have already lost their activity due to former gravitational compression of their current surface material.}

\keywords{Comets: general, Methods: analytical, Solid state: volatile}

\maketitle

\section{Introduction}\label{Introduction}
In recent years, observations of small bodies in the solar system have revealed the presence of low albedo asteroids in cometary like orbits \citep[ACOs;][]{Fernandezetal2001, Fernandezetal2005,Licandroetal2008}. In comparison to comets, ACOs are inactive small bodies, but with similar orbital parameters \citep{Fernandezetal2013,Kimetal2014,Belton2014}.
\par
\begin{figure}[t!]
    \centering
    \includegraphics[angle=180,width=1\columnwidth]{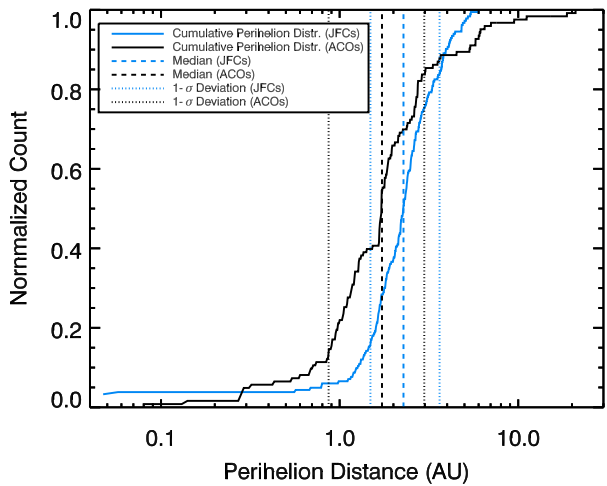}
    \caption{Cumulative perihelion distributions of the ACOs (black curve) and of the JFCs (blue curve). The median values of the two distributions are $1.73^{+1.25}_{-0.86} \, \mathrm{AU}$ (ACOs) and $2.28^{+1.35}_{-0.79} \, \mathrm{AU}$ (JFCs), respectively. For comparison, the mean values are $2.41^{+0.57}_{-1.54} \, \mathrm{AU}$ (ACOs) and $2.42^{+1.21}_{-0.93} \, \mathrm{AU}$ (JFCs). The errors denote one standard deviation of the perihelion distances. The data were taken from the JPL Small-Body Database and the work by \citet[][their Table 1]{Kimetal2014}.}
    \label{fig_orbits_cum}
\end{figure}

JFCs and ACOs possess a relatively similar cumulative size distribution between $\sim1\,\mathrm{km}$ and $\sim2\,\mathrm{km}$ in radius with power law indices of $1.1$ and $1.0$, respectively \citep[see Fig. 3a in][]{Kimetal2014}. However, for larger sizes ($\gtrsim2\,\mathrm{km}$ in radius), the two cumulative size distributions show a significant discrepancy in slope ($1.9$ and $1.1$, respectively). The similarity of the two cumulative size distributions in the small-size range indicates that ACOs and JFCs may belong to the same population of primitive bodies. Another argument for this hypothesis is given by the similarity between the Tisserand parameter of ACOs and JFCs \citep[$T_J < 3$;][]{LevisonDuncan1997, Kimetal2014}. Furthermore, ACOs and JFCs possess similar distributions of their orbital elements and, thus, of their perihelion distances (see Fig. \ref{fig_orbits_cum}), which is the most important orbital parameter when discussing comet activity. Most objects are on orbits between $\sim 1\,\mathrm{AU}$ and $\sim 5 \,\mathrm{AU}$\footnote{We used the JPL Small-Body Database and the work by \citet[][their Table 1]{Kimetal2014} to compile the orbital elements of the JFCs and ACOs.}. An analysis of the orbits shows that the mean perihelion distance of both families is almost identical ($2.41^{+0.57}_{-1.54} \, \mathrm{AU}$ for ACOs and $2.42^{+1.21}_{-0.93} \, \mathrm{AU}$ for JFCs; the errors denote one standard deviation of the perihelion distances), whereas the median of the perihelion distances of the ACOs is slightly shifted towards shorter distances ($1.73^{+1.25}_{-0.86} \, \mathrm{AU}$ for ACOs and $2.28^{+1.35}_{-0.79} \, \mathrm{AU}$ for JFCs; see dashed black and blue line in Fig. \ref{fig_orbits_cum}). The similarity of the perihelion distributions is a further argument that JFCs and ACOs belong to the same population of primitive bodies.
\par
If JFCs and ACOs stem from the same population of primitive bodies, they must have formed under the same physical conditions in the protoplanetary disc. Recently, \citet{Skorov2012} and \citet{Blum2014} have suggested that comets formed by the gravitational collapse of a pebble cloud composed of mm- to dm-sized dust and ice aggregates \citep{Johansen2007}, because only in this process do the bodies possess sufficiently low tensile strengths to allow for gas-driven dust activity \citep{Skorov2012,Blum2014,Blum2015}. Model calculations have shown that the collision speeds during the collapse are low enough to not destroy the collapsing aggregates during collisons \citep{WahlbergJoghansen2014, Loreketal2015}.
\par
Hydrostatic compression of the dust aggregate that comprise the bodies formed by gravitational instability can enduringly enhance the tensile strength between the aggregates by a memory effect (see Sect. \ref{The memory effect of dust aggregate layers} for details). This effect was first investigated for granular materials by \citet{Tomas2004}. Later, \citet{Blum2014} measured the memory effect of dust-aggregate layers and found that $2.7 \, \%$ of the applied compression strain is remembered by the dust-aggregate layers as tensile strength.
\par
Applied to a cometary nucleus, the compression inside the body is gravitational (or hydrostatic) in nature and causes a strengthening of the material. When a comet loses non-gaseous material from its surface due to the outgassing of its volatile constituents, even deeper layers are exposed. With progression of the erosion front, the tensile strength of the surface material increases. Surface gravity is usually negligible compared to van der Waals attraction in the case of km-sized objects. This increase in tensile strength can lead to inactivity (in terms of dust emission), if the momentum transfer by the out-flowing gas molecules to the dusty surface material is no longer sufficient to overcome the cohesion among the dust pebbles. However, evaporating gas can still diffuse through the porous surface material as long as the heat wave can reach the volatile constituents.
\par
\begin{figure}[t!]
    \centering
    \includegraphics[angle=180,width=1\columnwidth]{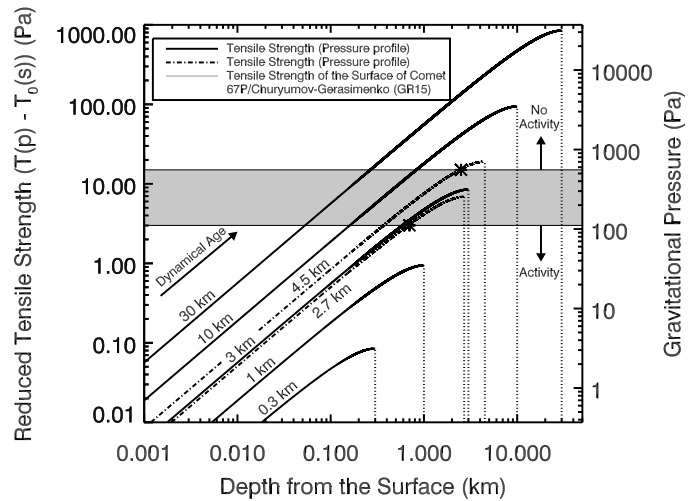}
    \caption{Memory-effect-induced increase in tensile strength (left y-axis) and hydrostatic compression (right y-axis) of the dust aggregates in a primitive body formed by gravitational collapse as a function of depth measured from the surface. The calculations were performed for five different initial radii of $0.3 \, \mathrm{km}$, $1 \, \mathrm{km}$, $3 \, \mathrm{km}$, $10 \, \mathrm{km}$ and $30 \, \mathrm{km}$, respectively (solid curves). The tensile strength of the dust aggregates is derived by the fact that $2.7\,\%$ of the applied compression strain is remembered by the dust aggregates as tensile strength. For comparison, the tensile strength of the surface material of comet 67P/Churyumov-Gerasimenko ($3 \, \mathrm{Pa}$ - $15 \, \mathrm{Pa}$) is shown by the grey area \citep[][]{Groussinetal2015}. To explain the presence of dust-active objects with $2 \, \mathrm{km}$ in radius and a surface tensile strength equal to the values measured for comet 67P/Churyumov-Gerasimenko, initial radii between $2.7 \, \mathrm{km}$ and $4.5 \, \mathrm{km}$ are required (dash-dotted curves; see main text for details). Objects below the grey area can always be active, while inactive bodies are located above. When dust removal from the surface takes place, the bodies develop along the line labelled dynamical age.}
    \label{intro}
\end{figure}

Figure \ref{intro} pictures the increase in hydrostatic compression, hence of the tensile strength, as the memory effect with increasing depth from the surface. Inside a homogenous body, the hydrostatic pressure as a function of depth, $z$, measured from the surface of the nucleus, is given by,
\begin{equation}
  p(z,R) \,  = \, \frac{2}{3} \, \pi \, \rho^2 \, G \, \left[ \, R^2 \, - \, \left( \, R \, - \, z \, \right)^2 \right] \, \mathrm{.}
\label{eq_0.1}
\end{equation}
Here, $\rho = 500 \, \mathrm{kg m^{-3}}$ is the assumed density of the body with radius $R$, and $G$ is the gravitational constant. The solid lines in Fig. \ref{intro} show the gravitational compression (right y-axis) as a function of depth for bodies with radii between $0.3 \, \mathrm{km}$ and $30 \, \mathrm{km}$. The vertical dotted lines denote the hydrostatic pressures in the body centres ($z = R$). Owing to the memory effect of dust-aggregate layers (see Sect. \ref{The memory effect of dust aggregate layers} for details), $2.7 \, \%$ of the applied gravitational compression is remembered by the dust-aggregate layers as tensile strength (left y-axis in Fig. \ref{intro}).
\par
Gas-driven dust activity is only possible if the internal pressure build-up due to the evaporating volatiles is sufficient to overcome the local tensile strength of the surface material. The Osiris camera onboard the Rosetta spacecraft provides the possibility to directly measure the tensile strength of comet 67P/Churyumov-Gerasimenko by calculating the force required to cause a collapse of overhangs \citep[][see their Eqs. 1 and 2, as well as their Figs. 9 - 11]{Groussinetal2015}. We used the tensile strengths ($3 \, - \, 15 \,\mathrm{Pa}$; see gray area in Fig. \ref{intro}) measured by \citet{Groussinetal2015} to derive how much material can be lost until the body becomes inactive. As long as the body is below the greay area in Fig. \ref{intro}, gas-driven dust activity is possible and we classify the respective object as an active JFC. If, however, the body shrinks in size such that it falls above this line, outgassing of the volatiles is still possible for some time, but the gas pressure is not strong enough to break the strengthened contacts between the aggregates. Thus, bodies above the grey area are to be categorized as inactive comets (known as ACOs). As can also be seen from Fig. \ref{intro}, larger bodies reach the transition between active and inactive comets faster (i.e., at a younger dynamical age, which is proportional to the number of perihelion passages and, therewith, with the thickness of the eroded material) than smaller objects, owing to their stronger hydrostatic compression. Very small objects (see, e.g., the $0.3 \, \mathrm{km}$ case in Fig. \ref{intro}) can remain as active comets during their entire lifetime (i.e., they never reach the grey area). In comparison to the larger objects, which will become inactive and remain as ACOs, the smaller bodies will be fully eroded until they vanish.
\par
This effect can be seen when comparing the cumulative size distribution of the two families. Larger bodies are lost from the population of the JFCs due to the vanishing activity and are then classified as ACOs. Therewith, the slope of cumulative size distribution of the JFCs should increase with time, whereas the slope of the cumulative size distribution of the ACOs should decrease with time. A significant difference is already visible for sizes larger than $\sim2 \, \mathrm{km}$ in radius \citep[see Fig. 3a in][]{Kimetal2014}.
\par
If we assume that the present-day radius of $2 \, \mathrm{km}$ determines the boundary between active and inactive objects and that the surface tensile strength of these objects is equal to the tensile strength measured for the surface of comet 67P/Churyumov-Gerasimenko \citep[$3 \, \mathrm{Pa}$ - $15 \, \mathrm{Pa}$][]{Groussinetal2015}, we can calculate their initial radii to fall between $2.7 \, \mathrm{km}$ and $4.5 \, \mathrm{km}$ (see dash-dotted curves in Fig. \ref{intro}). This comparison is accompanied by the implicit assumption that comet 67P/Churyumov-Gerasimenko is directly situated at the boundary between active and inactive objects at its present size.
\par
Bodies smaller than the critical radius of $2 \, \mathrm{km}$ will never become inactive due to their low tensile strength throughout their volumes (see Fig. \ref{intro}). Thus, no alteration of the two cumulative size distributions should occur for sizes smaller than $2 \, \mathrm{km}$ \citep[i.e., the cumulative size distributions of the ACOs and the JFCs possess a similar slope in the small size regime; see Fig. 3a in][]{Kimetal2014}, whereas above this size, significant difference in the two cumulative distributions should be observed.
\par
In this work, we formulate the hypothesis that JFCs and ACOs belong to the same population of primitive bodies. During their evolution, some objects lost their activity due to erosion of the surface layers and the resulting increase in tensile strength of the new surface material due to the memory effect of dust-aggregate layers (see Sect. \ref{The memory effect of dust aggregate layers}). The inactive comets are observed today as ACOs, while their active counterparts are known as JFCs. In Sect. \ref{Discussion}, we discuss why JFCs are active, while ACOs are inactive. Finally, the main findings of this work are summarized in Sect. \ref{Summary}.

\section{The memory effect of dust-aggregate layers}\label{The memory effect of dust aggregate layers}
Although there is strong evidence that ACOs and JFCs belong to the same original population of primitive bodies, the two main differences between these two families are that (i) ACOs extend to much larger sizes than JFCs and (ii) ACOs are non-active, whereas JFCs exhibit dust and gas activity. The memory effect described hereafter offers a natural explanation for these two differences. As can be seen in Fig. \ref{intro}, objects formed by gravitational instability of dust and ice aggregates and born larger can only lose a much thinner layer with low tensile strength before being rendered passive.
\par
The memory effect of granular matter was first experimentally studied by \citet{Tomas2004} for cohesive powders (e.g., limestone) and is being described simply as an inherent tensile strength proportional to the compressive strain applied over a period of time. However, the model derived by \citet{Tomas2004} is not directly applicable to predicting the tensile strength of dust-aggregate layers after hydrostatic compression, because it describes the behaviour of homogeneous granular matter with an intrinsically much greater tensile strength of several kPa \citep{Blum2006}, whereas cometary activity requires the tensile strength to not exceed $\sim 1\,\mathrm{Pa}$, which is only compatible with the formation of cometesimals by gravitational instability \citep{Skorov2012,Blum2014}.
\par
\citet{Blum2014} experimentally investigated the memory effect of mm-sized dust-aggregate layers, consisting of $\mu$m-sized $\mathrm{SiO_2}$ grains. They placed several layers of dust aggregates of a narrow size distribution onto a mesh inside a partially evacuated glass tube. Then, the samples were uniaxially compressed by a defined gas pressure differential for a duration of two minutes \citep[it was shown by different calibration experiments that the memory effect saturates for compression times longer than $100 \, \mathrm{s}$; see Fig. 6 in ][]{Blum2014}. After compression, the gas flow was inverted so that a tensile strain was applied to the pre-compressed dust-aggregate layers. Whenever this tensile strain was above the tensile strength of the sample, the specimen broke apart, which was recorded with a digital video camera. From the extrapolation of the tensile-strength data as a function of the applied compressive stress to zero compression, \citet{Blum2014} determined the tensile strengths of uncompressed dust-aggregate layers (their Figs. 7 and 8).
\par
Another result of these experiments was that the increase in tensile strength with applied compression, i.e., the memory effect, is proportional to the applied stress and seems to be independent of the dust-aggregate size. Although \citet{Blum2014} performed measurements for only two different dust-aggregate sizes that were a mere factor of two apart, we here assume that the memory effect does not depend on aggregate size.
\par
\begin{figure}[t!]
    \centering
    \includegraphics[angle=0,width=1\columnwidth]{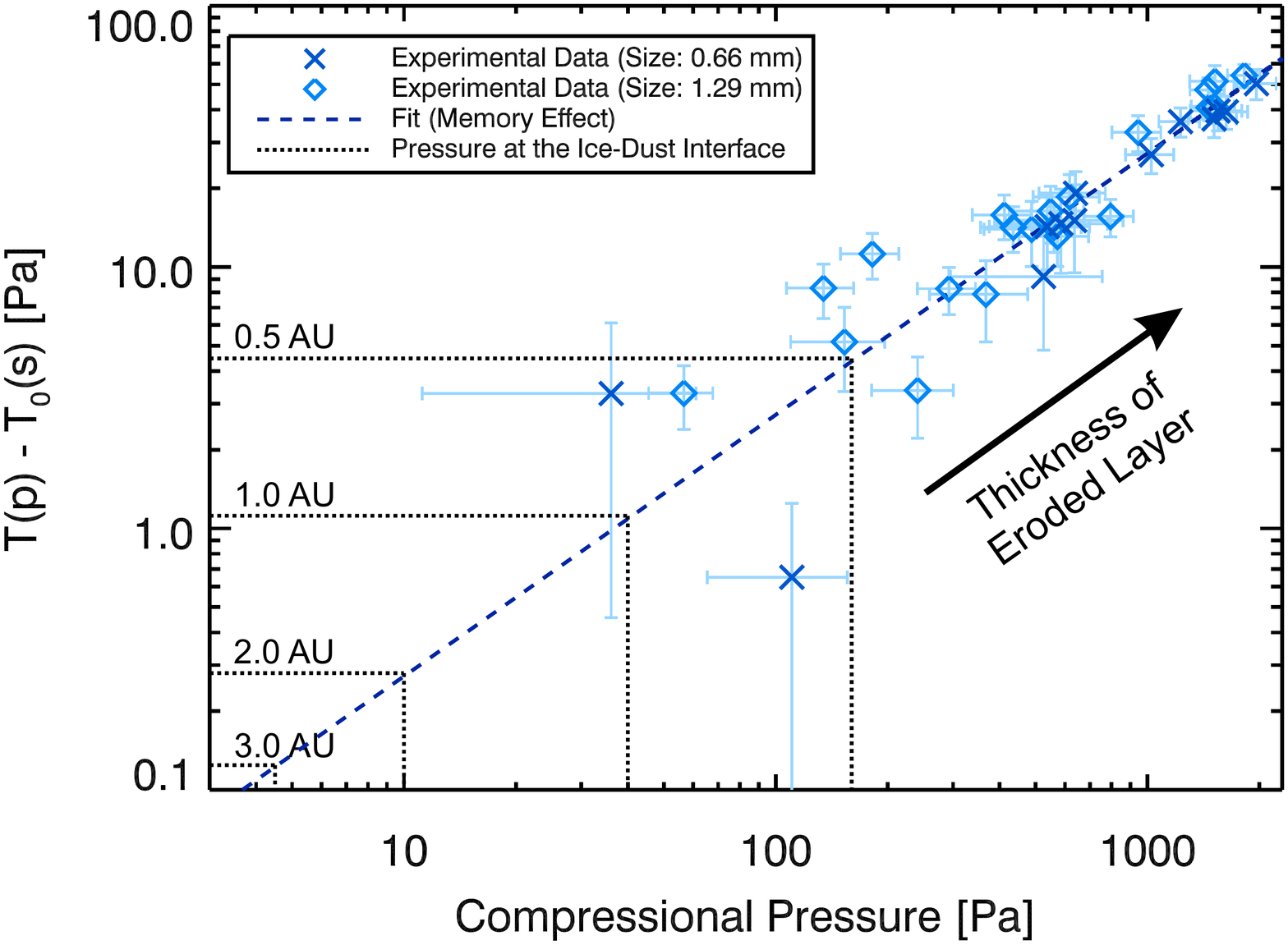}
    \caption{Tensile strength acquired through the memory effect, $T(p)\,-\,T_0(s)$, as a function of the compression $p$ experienced by dust-aggregate layers, as measured by \citet{Blum2014}. The blue crosses and diamonds represent dust-aggregate radii of $(0.66\pm0.14)\, \mathrm{mm}$ and $(1.29\pm0.29)\, \mathrm{mm}$, respectively. A linear fit to the data (blue dashed line) visualises the memory effect. For comparison, the dotted horizontal lines show the typical CO gas pressure at the ice-dust interface, i.e., the boundary between the covering non-volatile material and the ices, of a cometary nucleus derived for different heliocentric distances by using the model developed by \citet{Gundlachetal2015}. If the gas pressure at the ice-dust interface exceeds the tensile strength of the dust-aggregate layer, gas-driven dust activity is possible.}
    \label{fig_tensile_strength_all}
    \end{figure}

Here, we reconsider the data of \citet{Blum2014} for the memory effect and plot in Fig. \ref{fig_tensile_strength_all} the acquired tensile strength, namely $T(p)\,-\,T_0(s)$, as a function of the compressive stress $p$. Here, $T(p)$ and $T_0(s)$ are the tensile strength of the samples measured after applying a compressional stress $p$ and the tensile strength of the uncompressed samples, respectively, with $s$ being the radius of the dust aggregates. The symbols denote the two different dust-aggregate radii, i.e., $(0.66\pm0.14)\, \mathrm{mm}$ (crosses) and $(1.29\pm0.29)\, \mathrm{mm}$ (diamonds). As can be seen in Fig. \ref{fig_tensile_strength_all}, the two different aggregate sizes exhibit the same memory effect. A linear fit to the two data sets results in a slope of $(2.7\pm0.1) \times 10^{-2}$, which means that the dust-aggregate layer remembers $2.7 \, \%$ of the applied compressive stress as tensile strength. This value was used in Fig. \ref{intro} for deriving the tensile strength (left y-axis) from the hydrostatic pressure (right y-axis). The large scatter in the data towards lower values of the tensile strength results from the difficulty of measuring such low strengths, because the gravitational pressure of a single layer of mm-sized dust aggregates is of the order of $6 \, \mathrm{Pa}$. For the limiting case of zero compression, the memory effect vanishes. It is important to note that the tensile strength of the uncompressed samples, $T_0(s)$, depends on the size of the aggregates, while the increase in the tensile strength due to hydrostatic compression seems to be independent of the dust-aggregate size.
\par
For typical gravitational pressures for km-sized objects (e.g., between $\sim 10\,\mathrm{Pa}$ and $\sim 100\,\mathrm{Pa}$; see right y-axis of Fig. \ref{intro}), the increased tensile strength caused by the memory effect by far exceeds the tensile strength without compression. \citet{Skorov2012} calculated the intrinsic tensile strength of the contacts between dust aggregates of $35 \, \%$ volume filling factor due to van der Waals attractions among their monomer grains and got $T_0(s)=0.64 \, (s \,/ \, 1 \, \mathrm{mm})^{-2/3} \, \mathrm{Pa}$. Thus, for dust aggregates with radii of $s = 1 \, \mathrm{mm}$, $s = 1 \, \mathrm{cm}$, and $s = 1 \, \mathrm{dm}$, the intrinsic tensile strengths are $T_0 = 0.64 \, \mathrm{Pa}$, $T_0 = 0.14 \, \mathrm{Pa}$, and $T_0 = 0.03 \, \mathrm{Pa}$, respectively, which is only significant for the smallest, mm-sized, dust aggregates. This implies that for typical hydrostatic pressures in the interior of km-sized bodies, the size dependency of the tensile strength can be neglected.
\par
For comparison, typical gas pressures at the surface of a cometary nucleus obtained by the evaporation of of CO ice at different heliocentric distances ($0.5 \, \mathrm{AU}$, $1 \, \mathrm{AU}$, $2 \, \mathrm{AU}$, and $3 \, \mathrm{AU}$) are shown by the horizontal dotted lines in Fig. \ref{fig_tensile_strength_all} \citep[see][for details of the gas pressures calculations]{Gundlachetal2015}. If this gas pressure at the ice-dust interface exceeds the tensile strength of the overlying dust-aggregate layer, gas-driven dust activity is generally possible. For example, a dust-aggregate layer that has been hydrostatically compressed by $160 \, \mathrm{Pa}$ through the previously overlying material, requires a distance to the Sun of $0.5 \, \mathrm{AU}$, or less, to release the surface material by the outgassing of CO ice. A stronger gravitational compression of the surface material requires a closer distance to the Sun to allow for dust activity. If the material is less compressed, activity is also possible at larger heliocentric distances (see the other dotted lines in Fig. \ref{fig_tensile_strength_all}).

\section{Why are JFCs active and ACOs not?}\label{Discussion}
Here, we summarize four possible explanations for the fact that JFCs are active, while ACOs show no gas-driven dust activity.
\begin{enumerate}
\item JFCs are dynamically younger (fewer perihelion passages), which means that they have lost less surface material and have retained their activity. In this case, JFCs can in principle be arbitrarily large. However, larger objects will reach the inactivity threshold faster (i.e., less surface material must be eroded to reach inactivity; see Fig. \ref{intro}).
\item JFCs and ACOs have the same dynamical age, but JFCs possess larger perihelion distances than the ACOs (see Sect. \ref{Introduction} and Fig. \ref{fig_orbits_cum}). The closer perihelion distance of the ACOs leads to a faster erosion of their surfaces, which implies that the ACOs become inactive earlier than JFCs (although both populations may have experienced the same number of perihelion passages).
\item Larger comets become inactive faster than the smaller objects (see Fig. \ref{intro}). Thus, the cumulative size distribution of the JFCs is depleted for sizes $> 2\, \mathrm{km}$ in radius, in comparison to the cumulative size distribution of the ACOs \citep[see Fig. 3a in][]{Kimetal2014}. Comets smaller than $2\, \mathrm{km}$ in radius cannot lose their activity due to the low hydrostatic compression of the material.
\item ACOs are formed without icy materials, whereas JFCs are formed with a significant amount of volatile constituents.
\end{enumerate}
All four arguments alone, or a combination of them, are able to explain the activity of the JFCs and the inactivity of the ACOs. However, there is only observational support for the third point, namely that larger comets will become inactive first. This argument is corroborated by the fact that the activity of JFCs decreases with increasing size. \citet{Tancredietal2006} show that the fraction of active surface area of JFCs is relatively large (mostly $> 50 \, \%$) for bodies with radii $< 1\,-\,2\,\mathrm{km}$, whereas it decreases to $0\,-\,40 \, \%$ for objects with radii $> 2 \, \mathrm{km}$.
\par
However, what does this inactivity of the ACOs mean? Firstly, after reaching the transition between the JFCs and the ACOs (see grey area in Fig. \ref{intro}), the gas pressure is no longer sufficient to overcome the tensile strength of the new surface material, which implies that ejection of dust aggregates by gas drag is no longer possible. In this stadium, the comet is still active in terms of gas activity, i.e., the gas molecules can still diffuse through the porous surface material, but is inactive in terms of dust activity, i.e., dust coma and dust tail are no longer visible. Because the loss of dust is no longer possible, the sublimation front retreats farther away from the surface and the non-volatile dust layer grows. Thus, the thermal energy that reaches the evaporating ice front and, therewith, the reachable gas pressure decreases with time and the comet will gradually become inactive in terms of gas emission.
\par
The only two possibilities for reactivating a dormant comet is to increase the energy input by decreasing the perihelion distance, or to sublimate a more volatile component compared to the icy constituent that was responsible for the activity before.

\section{Summary}\label{Summary}
This work was motivated by the question of why JFCs are active, whereas ACOs show no gas-driven dust inactivity. To answer this question, we assumed that JFCs and ACOs belong to the same population of primitive bodies and that these objects have formed by the gravitational collapse of pebble clouds composed of mm- to dm-sized dust aggregates. Surfaces of bodies formed by this process possess an ultra-low tensile strength, which is required to explain gas-driven dust activity of comets. We showed in Fig. \ref{intro} that the hydrostatic compression inside these bodies leads to an increase in the tensile strength of the material caused by the memory effect (see Sect. \ref{The memory effect of dust aggregate layers}; dust-aggregate layers remember $2.7 \, \%$ of the applied compression as increased tensile strength).
\par
Active objects erode their surface layers by gas-driven dust activity as long as the gas pressure is sufficient to overcome the tensile strength of the material. With ongoing erosion of the object, deeper and deeper layers are exposed. With progression of the erosion front, the tensile strength of the surface material increases. For larger bodies, this increase in tensile strength can lead to inactivity if the gas pressure can no longer compete with the cohesion of the material anymore. At this point, an active JFC turns into an inactive body, i.e., an ACO.
\par
Because the cumulative size distributions of the JFCs and the ACOs start to show significant differences for sizes larger than $2 \, \mathrm{km}$ in radius \citep[see Fig. 3a in][]{Kimetal2014}, we used in-situ measurements of the tensile strength of the surface material of comet 67P/Churyumov-Gerasimenko \citep[][]{Groussinetal2015} to estimate the initial sizes of objects with a present-day radius of $2 \, \mathrm{km}$  (i.e., $2.7 \, \mathrm{km}$ and $4.5 \, \mathrm{km}$; see Fig. \ref{intro}). The current size of $2 \, \mathrm{km}$ seems to be a critical size above which the objects start to become inactive, which leads to a depletion of JFCs in this size regime; i.e., the slope of the cumulative size distribution of the JFCs is steeper compared to the cumulative size distribution of the ACOs. This depletion of active objects in the large-size regime is caused by a faster termination of gas-driven dust activity for larger bodies owing to the hydrostatic compression of the material. Objects smaller than $2 \, \mathrm{km}$ in radius cannot become inactive and are still present as JFCs. In the future, they will become fully eroded. Applied to comet  67P/Churyumov-Gerasimenko, this means that the larger lobe with a present radius\footnote{\url{http://www.esa.int/spaceinimages/Images/2015/01/Comet_vital_statistics}} of $\sim 2 \, \mathrm{km}$ could once have been as large as $2.7-4.5 \, \mathrm{km}$ and can still be active today.
\par
Thus, the question why JFCs are active while ACOs are inactive can be answered by four different scenarios (see Sect. \ref{Discussion}). Firstly, JFCs have experienced fewer perihelion passages than ACOs and have therefore lost less material (JFCs are dynamically younger). Secondly, JFCs and ACOs have the same dynamical age, but JFCs possess slightly larger perihelion distances than ACOs (see Fig. \ref{fig_orbits_cum}) so that the latter age on shorter timescales. Thirdly, JFCs are generally smaller than ACOs. In this case, the memory effect is less severe for the former, which implies that the larger objects will become inactive after fewer perihelion passages than initially smaller bodies on the same orbit. Fourthly, ACOs are born as asteroids (i.e., without a significant amount of icy materials) and JFCs are born as comets.

\bibliographystyle{aa}

\bibliography{bib}

\end{document}